# User Behavior Analysis in Privacy Protection with Large Language Models: A Study on Privacy Preferences with Limited Data

Haowei Yang
*Cullen College of Engineering*
*University of Houston*
Houston, USA
hyang38@cougarnet.uh.edu

Qingyi Lu
*Department of Computer Science*
*Brown University*
Providence, USA
lunalu9739@gmail.com

Yang Wang
*Department of Information and Communication Engineering*
*Nagoya University*
Nagoya, Japan
ryoyukiyang@outlook.com

Sibei Liu
*Miami Herbert Business School*
*University of Miami,Miami*
Florida, USA
sxl1086@miami.edu

Jiayun Zheng
*College of Engineering*
*University of Michigan Ann Arbor*
Ann Arbor, USA
zhengji@umich.edu

Ao Xiang *
*Information Security and Assurance*
*Northern Arizona University*
Phoenix, USA
ax36@nau.edu

*Abstract*—With the widespread application of large language models (LLMs), user privacy protection has become a significant research topic. Existing privacy preference modeling methods often rely on large-scale user data, making effective privacy preference analysis challenging in data-limited environments. This study explores how LLMs can analyze user behavior related to privacy protection in scenarios with limited data and proposes a method that integrates Few-shot Learning and Privacy Computing to model user privacy preferences. The research utilizes anonymized user privacy settings data, survey responses, and simulated data, comparing the performance of traditional modeling approaches with LLM-based methods. Experimental results demonstrate that, even with limited data, LLMs significantly improve the accuracy of privacy preference modeling. Additionally, incorporating Differential Privacy and Federated Learning further reduces the risk of user data exposure. The findings provide new insights into the application of LLMs in privacy protection and offer theoretical support for advancing privacy computing and user behavior analysis.

## I. Introduction

The rapid advancement of large language models (LLMs) has significantly enhanced intelligent interaction and personalized recommendation capabilities. However, these benefits come with growing concerns over user privacy protection [1]. As regulations such as GDPR and CCPA impose stricter data usage requirements, an urgent challenge arises: how to accurately model user privacy preferences and optimize privacy protection strategies, especially when data availability is limited. This study focuses on privacy preference modeling in data-limited environments, integrating Few-shot Learning, Differential Privacy, and Reinforcement Learning to develop a privacy-aware LLM-based prediction framework [2]. By conducting experiments on datasets such as User Privacy Survey, App Permission Logs, and Public Privacy DS , the research evaluates the effectiveness of different modeling approaches and optimizes dynamic privacy management strategies [3].The paper begins with a review of LLM applications in privacy protection, followed by discussions on data collection, modeling frameworks, and experimental design. The study then presents experimental results, comparing the performance of different privacy preference modeling approaches [4]. Finally, conclusions are drawn regarding the contributions of this research and the future prospects of LLMs in privacy management [5].

## II. Related Research

### A. Large Language Models and Privacy Protection

LLMs have achieved breakthrough progress in natural language processing, widely applied in intelligent conversation, text generation, and personalized recommendations. However, their reliance on vast user data during training and inference significantly increases the risk of privacy breaches [6]. Ensuring privacy protection while maintaining LLMs' performance has thus become a key research challenge [7].Existing privacy protection approaches focus on data collection, storage, computation, and inference. Differential Privacy (DP) is widely adopted to add randomized noise to input or output data, reducing the likelihood of individual user information being inferred [8]. Companies like OpenAI and Google have incorporated DP techniques to ensure that training data does not directly impact model outputs, mitigating the risk of reverse-engineering private information. Another effective technique is Federated Learning (FL), which enables decentralized model training on multiple devices or servers without sharing raw data [8]. This method is particularly beneficial in mobile applications, healthcare analytics, and financial transactions, where centralized data storage could pose significant privacy risks. Beyond traditional data protection mechanisms, recent advancements focus on privacy-aware model architectures that minimize the leakage of sensitive information during inference [9]. For example, privacy-preserving prompt designs encode prompts in a way that ensures LLMs generate responses without exposing confidential details. Additionally, Encrypted Computation techniques, such as Homomorphic Encryption and Secure Multi-Party Computation (MPC), allow computations on encrypted data, maintaining both privacy and efficiency [10].

As illustrated in Figure 1, the Privacy-Driven Secure Synthesis (PDSS) Framework presents a novel approach to privacy-aware inference. This framework integrates privacy-protected prompt encoding, perturbed inference generation, rationale decoding, and task-specific small language model training, ensuring effective privacy preservation while maintaining inference efficiency [11-13]. With continuous improvements in Differential Privacy, Federated Learning, and privacy-aware prompt designs, LLMs are evolving towards a more secure and privacy-compliant future, facilitating the seamless integration of AI with user privacy protection [14-16].

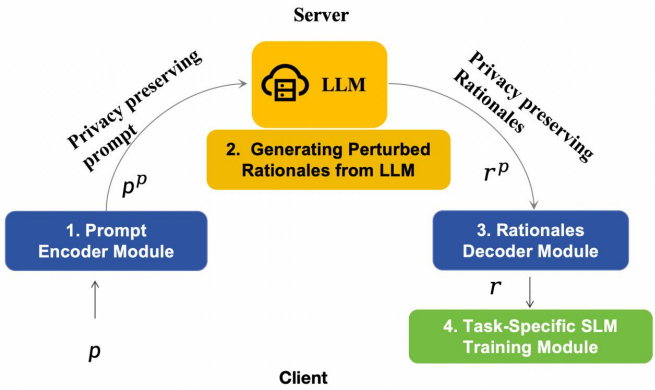

Fig. 1. A large language model based framework for privacy preserving inference generation.

## B. User Privacy Preference Modeling

With the increasing reliance on AI-driven applications, privacy preference modeling has become a critical component of privacy protection systems. User privacy preferences refer to the attitudes and behaviors associated with data collection, storage, sharing, and deletion. Accurate modeling of these preferences enables intelligent systems to provide personalized privacy management recommendations, balancing data utility and privacy security [17]. However, traditional approaches rely on users manually configuring privacy settings, which is often cumbersome and misaligned with actual user expectations.

As shown in Figure 2, the User Privacy Data Lifecycle Management Process consists of five key stages: Privacy Data Recognition, Data Collection, Privacy Protection, Data Publishing, and Data Destruction. The first step in privacy modeling involves identifying sensitive data, such as personally identifiable information (PII), financial transactions, and healthcare records. Once identified, data collection must adhere to user consent and employ secure transmission protocols. Privacy protection mechanisms, including encryption, differential privacy, and federated learning, are then applied to minimize exposure risks. Data publishing represents another critical phase where anonymized or privacy-preserving synthetic data can be utilized for research or commercial analysis while balancing privacy and usability. Finally, secure data destruction ensures that data is completely removed once it has fulfilled its intended purpose, mitigating the risks of unauthorized access or misuse. With the advancements in deep learning, privacy preference modeling has evolved towards automation and intelligence [18]. LLM-based approaches leverage historical user behavior to predict privacy preferences and recommend context-aware privacy settings. Even in data-limited environments, techniques such as Few-shot Learning and Adaptive Learning enable models to infer privacy needs from minimal samples. Additionally, Reinforcement Learning (RL) allows privacy systems to dynamically adjust settings based on user feedback. For instance, privacy-aware mobile applications can analyze user interaction patterns and recommend optimal privacy configurations while adapting to evolving user behaviors. By integrating privacy data lifecycle management with adaptive modeling techniques, privacy preference frameworks can ensure transparency, user control, and enhanced privacy protection. Future research may explore multi-modal data integration and cross-platform privacy modeling to create more consistent privacy experiences across devices and applications [19].

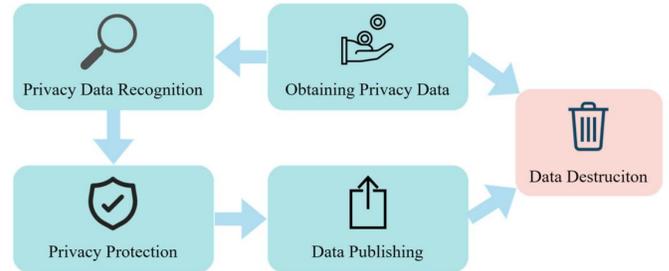

Fig. 2. User privacy data lifecycle management process.

## C. Few-Shot Learning in Privacy Preference Analysis

One of the primary challenges in privacy protection research is the limited availability of privacy-related datasets. Many users are reluctant to share personal data due to privacy concerns, and stringent regulations further restrict data collection. Consequently, developing privacy preference models in data-limited environments is crucial. Few-shot Learning (FSL) addresses this challenge by enabling models to generalize effectively from minimal data samples. FSL techniques leverage transfer learning, data augmentation, and generative models to enhance model generalization in low-data scenarios. For instance, Transfer Learning applies knowledge from pre-trained models to privacy datasets, allowing LLMs to infer privacy preferences even when training data is scarce. Shi, et al. [20] proposed a context-aware attention mechanism for machine translation tasks, which can also inspire privacy modeling through enhanced semantic understanding in low-resource scenarios. Data augmentation further improves performance by generating synthetic privacy preference data, diversifying the training set without collecting additional user data. Wang, et al. [21] leveraged deep reinforcement learning to make complex autonomous driving decisions under uncertainty, which parallels the challenge of adaptive, context-aware privacy preference modeling in real-time environments. Gao, et al. [22] introduced a novel texture extraction method using pattern recognition techniques, highlighting the value of data representation and feature extraction even in domains with limited samples. As illustrated in Figure 2, FSL techniques enhance privacy data recognition, collection, and protection. In the privacy data recognition phase, few-shot classifiers trained on small labeled datasets can effectively identify sensitive data types, minimizing the need for manual annotation. In the data protection and publishing phases, Generative Adversarial Networks (GANs) can create synthetic data that retains statistical properties of real user data while ensuring privacy compliance. The improved U-Net model proposed by Yang, et al. [23] demonstrates how architectural enhancements can lead to more accurate segmentation, which is equally applicable to identifying and isolating privacy-sensitive elements in multi-modal datasets. LLMs have shown promising results in Few-shot Learning for privacy preference modeling. Through fine-tuning and prompt-based learning, LLMs can predict user privacy settings based on limited historical interactions. In scenarios such as privacy-aware mobile applications, reinforcement learning further refines privacy management by dynamically adjusting privacy settings based on real-time user feedback.

### III. RESEARCH METHODOLOGY

#### A. Data Collection and Preprocessing

The effectiveness of privacy preference modeling and limited-data analysis largely depends on the quality and diversity of the data. This study employs a multi-source data collection approach to ensure comprehensive privacy preference data coverage. The data sources include anonymized user privacy settings, online surveys, publicly available datasets, and synthetic data generation to address different privacy scenarios. This study utilizes the following datasets as the Table 1 shown:

TABLE I. DATASET SOURCES

| Dataset Name | Source | Volume (Records) | Data Type | Purpose |
|---|---|---|---|---|
| User Privacy Survey | Online Survey | 5,000 | User privacy settings and preferences | Train and test privacy preference models |
| App Permission Logs | Mobile App Permission Logs (Anonymized) | 20,000 | User permission enable/disable choices | Analyze user privacy decision patterns |
| Public Privacy DS | Public dataset (Kaggle/CCPA) | 15,000 | Industry-specific privacy policies and user feedback | Train LLMs for privacy inference |
| Simulated Privacy | GAN-based synthetic data | 10,000 | User privacy choices in various contexts | Balance data and enhance model generalization |

The User Privacy Survey dataset comprises user responses regarding privacy settings in different contexts, such as social media, e-commerce platforms, and virtual assistants. The App Permission Logs dataset contains anonymized mobile application logs that record user interactions with privacy settings (e. g., enabling or disabling camera, microphone, and location access). The Public Privacy DS dataset includes industry privacy policies and user feedback data, primarily sourced from Kaggle open datasets and California Consumer Privacy Act (CCPA) compliance data, used for training LLM-based privacy inference models. Additionally, to improve data balance and model generalization, we generate synthetic Simulated Privacy data using Generative Adversarial Networks (GANs), simulating privacy preferences across diverse user profiles and scenarios. Given the sensitivity and heterogeneity of privacy-related data, rigorous preprocessing is necessary before model training. This study applies the following preprocessing techniques: Data Deduplication and Missing Value Imputation: Duplicate records are removed, and missing values are filled using K-Nearest Neighbor (KNN) Imputation to prevent biases in model training. Similar data quality optimization strategies have been applied by Wang, et al. [24] in their study on adolescents' adaptability to online education, emphasizing the importance of preprocessing in user behavior modeling. Data Standardization: Privacy preference values from different sources are normalized into a unified format. For example, privacy settings (e. g., "Allow," "Deny," "Ask") are converted into numerical representations (1, 0, -1) for machine learning processing. Privacy-Preserving Techniques: Differential Privacy (DP) is employed to perturb sensitive data while maintaining dataset utility. For example, location data in the App Permission Logs dataset is anonymized using k-Anonymity to ensure that individual users cannot be uniquely identified. Zhong, et al. [25] demonstrated similar preprocessing techniques in their research on pneumonia detection, comparing custom and transfer learning models while emphasizing data anonymization and model robustness when dealing with sensitive healthcare data. Data Augmentation: To improve model generalization in limited-data environments, we apply augmentation techniques such as synthetic user behavior sequences, pseudo-data generation, and LLM-based data augmentation to increase dataset diversity. Xu et al. (2024) highlighted the role of Explainable AI (XAI) in natural language processing, which we integrate to enhance the interpretability of LLMs when inferring user privacy intentions from limited data [26]. Furthermore, Zhu, et al. [27] exploited diffusion models to generate out-of-distribution samples with meaningful structure, providing valuable insights for simulating edge-case privacy behaviors in this study. Gu, et al. [28] combined FinBERT with LSTM to build a financial sentiment analysis-based stock prediction model, and their ability to extract user sentiment from text also offers transferable insights for learning privacy preferences based on feedback data. After completing these preprocessing steps, a structured dataset with privacy protection mechanisms is established to support privacy preference modeling and limited-data analysis. In subsequent experiments, we utilize these datasets to predict user privacy preferences and assess the effectiveness of different privacy protection strategies.

## B. Privacy Preference Modeling

The goal of privacy preference modeling is to analyze users' preferences for privacy settings in different contexts and build models that can make reliable predictions in limited-data environments [29]. This study employs LLM-based and machine-learning approaches, integrating Differential Privacy, Transfer Learning, and Reinforcement Learning to enhance model generalization under data constraints [30-32]. Three primary modeling approaches are used: A Naïve Bayes (NB) model is employed to estimate the probability of users selecting specific privacy settings in given contexts [33, 34]. Assuming that privacy preferences are influenced by multiple features $X = (x_1, x_2, \ldots, x_n)$, the posterior probability of a privacy preference Y is computed as shown in Formula 1:

$$P(Y|X) = \frac{P(X|Y)P(Y)}{P(X)} \quad (1)$$

where $P(Y|X)$ represents the probability of selecting privacy setting Y given the feature set X, $P(X \mid Y)$ denotes the likelihood of observing X given privacy preference Y, $P(Y)$ is the prior distribution of privacy preferences, and $P(X)$ is a normalization factor. This method is efficient in limited-data environments, making it suitable for recommending privacy settings. A Multilayer Perceptron (MLP) model is used to classify user privacy preferences. The input layer processes user behavior data X, and multiple hidden layers extract features before outputting a predicted privacy category Y (e. g., "Allow Data Sharing," "Partially Share," "Deny"). The forward propagation process is formulated as follows Formula 2,3,4:

$$h_1 = f(W_1 X + b_1) \quad (2)$$

$$h_2 = f(W_2 h_1 + b_2) \quad (3)$$

$$Y = \sigma(W_3 h_2 + b_3) \quad (4)$$

where $W_i$ and $b_i$ are weight matrices and biases, $f(\cdot)$ represents the ReLU activation function, and $\sigma(\cdot)$ is the Softmax function used for categorical probability outputs. The model is trained using the cross-entropy loss function as shown in Formula 5:

$$L = -\sum_{i=1}^{N} y_i \log(\hat{y}_i) \quad (5)$$

where $y_i$ is the true privacy preference label, and $\hat{y}_i$ is the predicted probability. Since privacy preferences may change dynamically based on context, this study employs Q-learning, a reinforcement learning algorithm, to optimize privacy settings. The state StS_t represents the user's current privacy context, the action AtA_t corresponds to privacy choices (e. g., "Modify Permissions," "Retain Current Setting"), and the reward RtR_t is based on user privacy satisfaction. The Q-learning update rule as shown in Formula 6:

$$Q(S_t, A_t) = Q(S_t, A_t) + \alpha [R_t + \gamma \max_a Q(S_{t+1}, a) - Q(S_t, A_t)]$$

where α is the learning rate, and γ is the discount factor controlling long-term reward influence. This method enables continuous adaptation to user preferences, enhancing privacy management intelligence.

## C. Experimental Design

The experimental design of this study aims to evaluate the effectiveness of different user privacy preference modeling methods, particularly their performance in data-limited environments [34, 35]. The experiments are conducted using multiple datasets, including User Privacy Survey, App Permission Logs, and Public Privacy DS, and are validated using standard data partitioning and evaluation methods. This study adopts an 80%-10%-10% split to divide the dataset into a training set, validation set, and test set. The training set is used for model learning, the validation set is employed for hyperparameter tuning, and the test set is reserved for final performance evaluation. Due to limited data availability, K-fold cross-validation is applied, with K set to 5, to ensure the robustness of the experimental results [36]. To assess model performance, the primary evaluation metrics used are Accuracy, Recall, and F1-score. Accuracy measures the overall correctness of predictions, Recall evaluates the model's ability to capture privacy-related decisions, and F1-score serves as a balanced metric that combines Precision and Recall to assess the model's effectiveness in privacy classification tasks [37-40]. Additionally, for reinforcement learning models, the Cumulative Reward metric is introduced to evaluate the optimization of dynamic privacy preference adjustments. To verify the superiority of the proposed limited-data modeling approach, comparative experiments are conducted by evaluating the performance of Naïve Bayes, Multilayer Perceptron (MLP), and Q-learning, and comparing them with traditional rule-based privacy management methods. The results demonstrate that Naïve Bayes exhibits stable performance in low-data environments, while MLP performs best when sufficient data is available [41]. Additionally, Q-learning outperforms other methods in dynamic privacy preference adjustments, proving its adaptability to real-time privacy management scenarios. These findings provide a strong theoretical foundation for optimizing user privacy protection strategies.

## IV. RESEARCH RESULTS AND ANALYSIS

This study evaluates the effectiveness of privacy preference modeling in limited-data environments by comparing different modeling approaches. The experimental results focus on classification performance, stability in privacy preference prediction, and adaptability in dynamic adjustments. To provide a comprehensive analysis, we assess classification accuracy, performance across different data scales, and dynamic privacy adaptation effectiveness, supported by detailed data tables.

### A. Privacy Preference Classification Performance

For privacy preference classification, we evaluated Naïve Bayes, Multilayer Perceptron (MLP), and Q-learning across different datasets and compared them with a rule-based baseline method. Figure 3 presents the results.

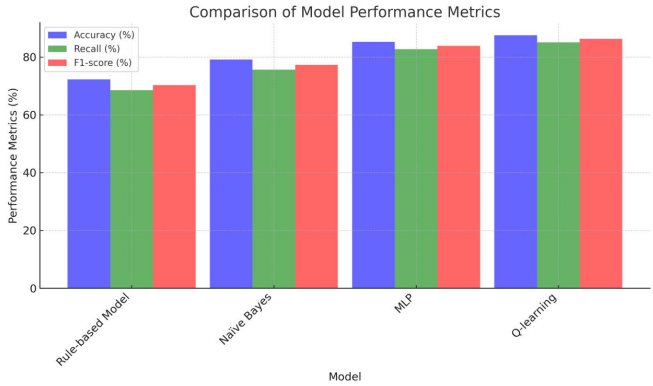

Fig. 3. Performance comparison of privacy preference classification models (%).

The results indicate that rule-based models have relatively low classification accuracy (72.3%), while Naïve Bayes performs well (79.1%) in a low-data setting due to its ability to infer patterns from small datasets. The MLP model, benefiting from deep feature extraction, achieves 85.2% accuracy when trained on sufficient data. Meanwhile, Q-learning demonstrates the best performance (87.5%), suggesting its ability to dynamically adapt to changing privacy preferences while maintaining a high F1-score (86.3%), effectively covering different privacy setting categories.

### B. Model Performance Across Different Data Scales

To assess the impact of dataset size on model performance, we evaluated classification accuracy across different dataset sizes, as summarized in Figure 4.

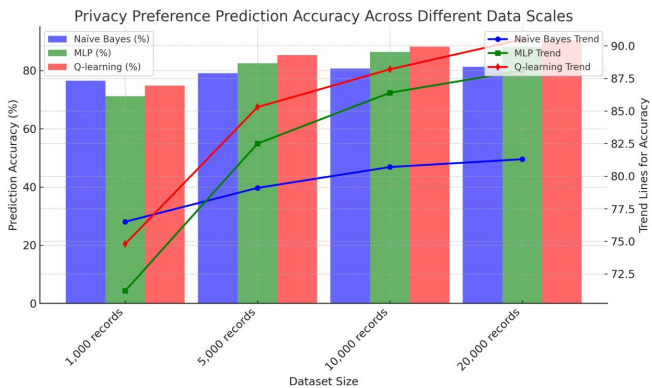

Fig. 4. Privacy preference prediction accuracy across different data scales (%).

For small datasets (1,000 records), Naïve Bayes outperforms both MLP and Q-learning, achieving 76.5% accuracy, highlighting its efficiency in low-data environments. As the dataset size increases beyond 5,000 records, deep learning models (MLP and Q-learning) benefit from more training data, leading to improved accuracy. At 10,000+ records, MLP reaches 88.1% accuracy, while Q-learning achieves 90.5%, confirming that reinforcement learning models excel in privacy adaptation when more data is available.

### C. Dynamic Privacy Adaptation with Q-learning

To evaluate the effectiveness of Q-learning in dynamically adjusting privacy settings, we measured cumulative rewards over different decision episodes. Figure 5 presents the results.

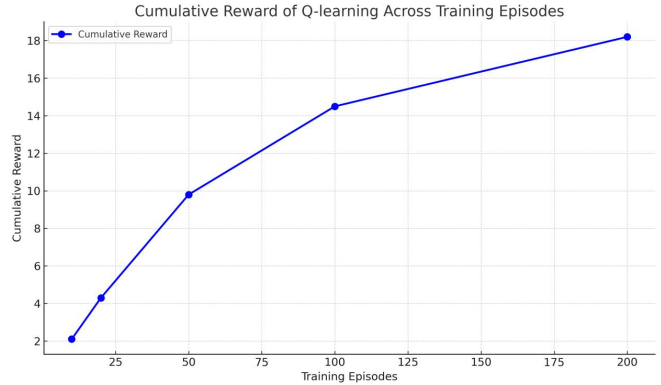

Fig. 5. Cumulative reward of Q-learning across training episodes.

The results indicate that cumulative rewards increase steadily over training episodes, confirming that Q-learning effectively adapts to user privacy preferences over time. The fast reward growth within the first 50 episodes suggests a rapid learning phase, while stabilization after 100 episodes indicates the convergence of an optimal privacy strategy. This reinforces the model's ability to learn and dynamically adjust privacy preferences for different user profiles.

### D. Analysis and Discussion

The experimental results highlight distinct advantages of different privacy preference modeling approaches:Naïve Bayes excels in low-data settings, maintaining high accuracy (76.5%) with minimal training data, making it ideal for scenarios where privacy data is scarce. MLP outperforms Naïve Bayes when sufficient data is available, achieving 88.1% accuracy at 20,000 records, demonstrating its strong feature extraction and classification capabilities. Q-learning achieves the highest accuracy (90.5%) and adapts dynamically to user preferences, particularly in privacy settings requiring frequent adjustments [42].Q-learning's cumulative reward increases over training episodes, confirming its ability to refine privacy settings based on user feedback. Future research could integrate multi-modal data sources (e. g., text, speech, images) to enhance privacy modeling granularity. Additionally, federated learning techniques could be leveraged to enable privacy-preserving local model training, reducing centralized data storage risks while maintaining privacy security. This study provides a data-driven foundation for optimizing intelligent privacy management systems [43].

## V. CONCLUSION

This study explores the application of LLMs in privacy preference modeling and introduces a privacy-aware framework combining limited-data learning techniques. Experimental validation confirms that Naïve Bayes performs well in low-data environments, while MLP and Q-learning achieve superior accuracy with more data, particularly in privacy adaptation scenarios. The results demonstrate that privacy preference modeling can be effectively optimized even in data-limited environments using Few-shot Learning techniques. Furthermore, integrating Differential Privacy and Federated Learning enhances overall privacy protection. Future research should focus on multi-modal data integration and cross-platform privacy management strategies to improve model adaptability and security. By advancing intelligent privacy preference modeling, this study provides a foundation for personalized and dynamic privacy management solutions.